\documentclass[11pt,twoside]{article}
\usepackage{asp2004}
\usepackage{psfig}
\usepackage{epsf}
\usepackage{graphics}
\usepackage{lscape}
\def\edcomment#1{\iffalse\marginpar{\raggedright\sl#1\/}\else\relax\fi}
\reversemarginpar

\begin{document}
\title{The relation between optical and X-ray variability in Seyfert galaxies}
\author{P. Uttley}
\affil{NASA Goddard Space Flight Center, Code 662, Greenbelt, MD 20815}

\begin{abstract}
Studying simultaneous optical and X-ray light curves of radio-quiet AGN
can help to probe the relationship between
very different physical components
- the cool, optically thick disk and hot, optically thin corona.
Here, we review the relationship between optical and X-ray variability in Seyfert galaxies,
which due to observing constraints was difficult to study for many years, but was given a huge boost
with the launch of the {\it RXTE} satellite in 1995. 
We summarise the diverse results of several monitoring campaigns,
which pose a challenge for standard theories relating optical and X-ray variability, with
sources showing either correlated optical and X-ray flux
variations, correlated optical flux and X-ray spectral variations, or no correlation at all.
We discuss possible explanations for these results, some of which may be explained using
a more standard AGN picture, while others may require additional components, such as the
2-phase accretion flows suggested to explain black hole X-ray binary behaviour.
\end{abstract}
\thispagestyle{plain}

\section{Introduction}
Temporal variability of emission across the electromagnetic spectrum is a key
characteristic of all AGN.  In Seyfert galaxies and radio-quiet quasars, different
components - optically thin corona and optically thick disk - are thought to
produce the continuum emission in X-ray and optical/UV
wavebands respectively\footnote{Since the optical and UV emission 
are thought to originate from the same optically thick blackbody component, we will 
refer to them interchangably when discussing variability.}.  Studying the relation
between variability in these different bands can provide important clues about 
how the disk and coronal emission are connected, and help answer the key question 
of what causes the variability in the first place.  

It has been known for many years that
variability is strongest and most rapid at the highest energies, in the X-ray band 
and above (e.g. see McHardy, these proceedings), and light-travel-time arguments
then imply that the X-rays originate close to the central black hole, where most of
the accretion energy is liberated.  Therefore, if the central engine also drives variability
in other wavebands, we might expect a correlation of some sort between X-ray and
optical emission.  For example, the X-rays might be reprocessed into optical photons by
heating the disk \citep{kro91}.  Alternatively, if optical photons provide the source
of `seed' photons for the X-ray Comptonisation process, the X-ray variability may track the 
optical variability.  In either case, a correlation between X-ray and optical
variations
would be expected, perhaps with measurable time-lags which indicate which is the
driving continuum band.  Even if the optical and X-ray emitting regions are physically
separate, a common variability mechanism might lead to correlated optical/X-ray 
variability, for example if underlying accretion rate variations drive the 
variability in both bands.

Perhaps surprisingly, it was not until the last decade of the last century, some twenty
years after the dawn of X-ray astronomy, that observations of simultaneous optical/X-ray
variability started to be made.  The difficulty in obtaining these observations lay primarily
with the inflexible scheduling of X-ray satellites, which are generally not designed
to carry out multiple, well-sampled monitoring observations required to study
simultaneous optical/X-ray variability.  Furthermore, until recently,
the difficulty of scheduling  large ground-based monitoring campaigns has made 
organising multiwavelength monitoring
campaigns an almost heroic task - as a glance at the lengths of author lists of early papers
in the subject will testify!  However, the situation improved in December 1995
with the launch of the Rossi X-ray Timing Explorer ({\it RXTE}), which has a rapid pointing 
capability and a scheduling ethos
that allows well-sampled X-ray monitoring campaigns to run parallel with 
long-term optical monitoring for weeks or even years.
In this review I will consider the results of the last decade or so of simultaneous 
X-ray/optical monitoring campaigns of Seyfert galaxies, obtained prior to {\it RXTE}
and in the {\it RXTE} era, which reveal a surprisingly
complex picture of the relation between optical and X-ray
variability in Seyfert galaxies.

\section{Before {\it RXTE}}
A seminal paper studying the relation between optical and X-ray variability
was published by \citet{don90} who, over the course of two 
nights, used simultaneous optical photometric observations and X-ray observations
with the Japanese {\it Ginga} satellite to monitor the simultaneous optical/X-ray
variability of  the Narrow Line Seyfert~1 galaxy NGC~4051.  The campaign
 revealed the physically interesting, if disappointing, result that the X-rays 
varied significantly
(by more than a factor 2) while the optical emission did not appear to vary at all (to within
photometric 1\% accuracy).  This result primarily showed that the optical and X-ray 
continua didn't originate from the same component, e.g. the same population of 
synchrotron-emitting electrons.  But the fact that the optical emission was much less
variable than the X-ray emission could still be reconciled with reprocessing models if
the reprocessor in NGC~4051 was very large (e.g. light-days), so that light travel 
time effects dilute the variability, or if only a small fraction of the optical emission is due
to reprocessed X-rays (as might be 
expected if internal heating in the disk dominates over the external heating by X-rays).

Although Seyferts show very weak optical variability on time-scales of days or less, 
on longer time-scales the amplitude of variability increases, as had been demonstrated since
the first AGN optical monitoring campaigns began several decades ago\footnote{Many of them
at the Crimean Astrophysical Observatory where this meeting was held}.  Therefore,
it was realised that longer-term monitoring in both optical and X-ray wavebands would
provide a better probe of the relation between the two bands than observations over just
a few days.  However, the difficulty with scheduling X-ray monitoring observations limited 
the number of such campaigns, which were primarily accomplished using {\it ROSAT} to
monitor the variability in the soft X-ray band.
The results of these campaigns were suggestive of an optical/X-ray correlation 
on time-scales of weeks to months in the Seyfert~1 galaxies NGC~5548 \citep{cla92} and
NGC~4151 \citep{ede96}.  However, since both correlations were dependent on only one
`event' (a rise or fall) in the light curve, it couldn't be proven conclusively that the
light curves were really correlated, or if the observed correlations were an artefact
of the `red-noise' nature of the variability\footnote{Red-noise light curves show 
temporally-correlated variability, i.e. one data point is correlated with the next in the 
time series,
and so normal cross-correlation statistics, which assume temporally-uncorrelated data, cannot be used
to estimate the significance of a correlation between light curves in two
energy bands.  Either an extensive data set must be 
obtained, much longer than time-scales present in the light curve 
(usually impossible with AGN), or Monte-Carlo simulations should be performed.}.  
Clearly, longer simultaneous light curves were needed, covering time-scales of years.

\section{The {\it RXTE} era}
With the launch of {\it RXTE}, it became possible to obtain well-sampled X-ray light
curves of AGN covering a very broad range of time-scales.  Many such light curves have
been obtained by campaigns to measure the broadband X-ray power spectra of
Seyfert galaxies, e.g. \citet{utt02,mar03} and see McHardy, these proceedings.
Fortunately a number of Seyfert galaxies with X-ray monitoring
have also been observed by various optical monitoring campaigns, revealing
clearly the relations between long-term optical and X-ray variability for the first time.
The picture revealed by these campaigns is diverse, and we concentrate here 
on several case studies of specific AGN,
which reveal the range of behaviour observed in the multiwavelength campaigns 
carried out so far.

\subsection{NGC~7469}
\begin{figure}[!ht]
\plotfiddle{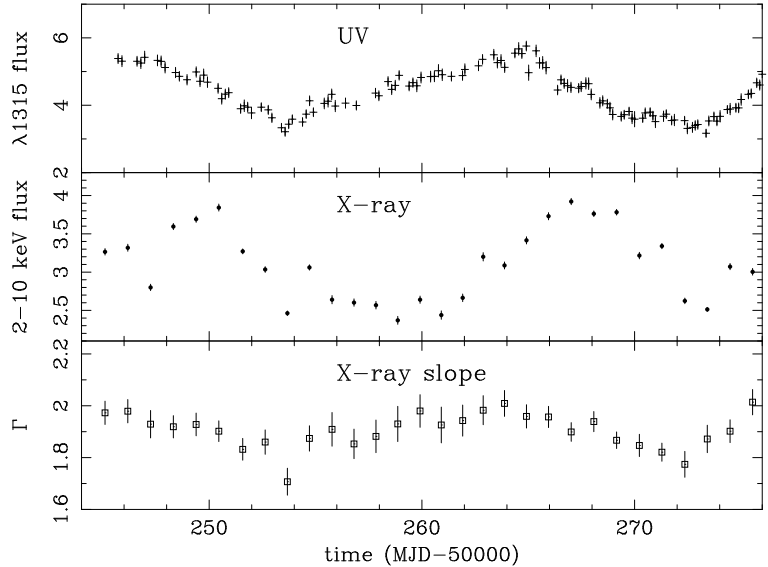}{8cm}{0}{60}{60}{-150}{-50}
\caption{NGC~7469 light curves of 1315\AA\ flux
($10^{-14}$~erg~cm$^{-2}$~s$^{-1}$~\AA$^{-1}$),
2-10~keV X-ray flux ($10^{-11}$~erg~cm$^{-2}$~s$^{-1}$) and X-ray spectral
photon index, $\Gamma$ (data taken from \citealt{nan00}).} \label{ngc7469lcs}
\end{figure}

The first extensive multiwavelength monitoring campaign in the {\it RXTE} era was 
carried out in 1996, with a month-long {\it RXTE} campaign to monitor
the Seyfert~1 galaxy NGC~7469,
simultaneous with UV observations by the International Ultraviolet Explorer ({\it IUE})
\citep{nan98}.  The surprising result of this campaign was that the X-ray flux was not 
correlated with the variable UV flux in any obvious way \citep{nan98}.  However,
a subsequent X-ray spectral analysis of the X-ray data suggested that the X-ray 
continuum slope and the UV flux were correlated (\citealt{nan00}, and see 
Fig.~\ref{ngc7469lcs}).  The reality of this correlation between X-ray spectral properties
and UV flux was subsequently confirmed using Monte Carlo simulations by
\citet{pet04}.

At first glance, the correlation between X-ray spectral slope and UV flux might be
explained if the UV luminosity dominates that of the corona, and UV
variability drives the X-ray variations through Comptonisation,
with the increases in UV flux acting to cool the X-ray emitting corona, steepening
the resulting X-ray spectrum.  However, in that case,
if the UV flux simply increases, one should expect
correlated increases in X-ray flux (as well as spectral steepening), which are not observed.  
Instead, \citet{pet04} explained the data in terms of a model where the 
coronal power is dominant and X-rays drive UV variations
through reprocessing, requiring almost all of the UV flux to be produced by X-ray heating 
(i.e. internal heating is negligible).  However, their model
also suggested that the X-ray (and resulting UV) variations are primarily driven by
changes in the geometry of the X-ray source, thus changing the fraction of UV
photons that are Comptonised and hence cool the corona.

\subsection{NGC~4051}
NGC~4051 which, as \citet{don90} showed, does not vary significantly in the optical on 
time-scales of hours, has been monitored extensively with {\it RXTE} since it was launched
(see M$^{\rm c}$Hardy, these proceedings, and \citealt{mch04}).  Long-term
optical monitoring by the International AGN Watch during the first three years 
of the {\it RXTE} campaign showed significant but weak ($<10$\% fractional rms) 
variability on long time-scales, which none-the-less appeared to be correlated with the
much greater amplitude ($>60$\% fractional rms) long-term X-ray variability \citep{pet00}.

On shorter time-scales, a 1.5 day observation using
the Optical Monitor (OM) on board {\it XMM-Newton}
revealed correlated UV/X-ray variability, with a fractional rms of a few per cent and the
smooth near-UV flux variations appearing to lag the rapid, large-amplitude
X-ray variations by about 0.2 days \citep{mas02}.  The data could be simply explained if the UV
variations were caused by reprocessing of X-rays in a ring $\sim0.14$~light-days 
from the X-ray source - similar
to the radius where near-UV emission is expected to peak in a standard accretion disk around
a $\sim10^{6}$~M$_{\odot}$ black hole (which is close to the black hole
mass estimated from reverberation mapping, \citealt{pet00}).  The small amplitude of
variability would suggest that rather constant
internal heating dominates the near-UV emission in this case\footnote{Presumably 
\citet{don90} did not see these short-term variations because
their observations only lasted a few hours at a time, and occured at longer 
wavelengths where variability amplitudes may be weaker.}.

However, the story for
NGC~4051 may not be so simple, as daily optical monitoring, together with
intensive {\it RXTE} monitoring suggested correlated variability, but with optical variations
{\it leading} X-rays by an average of about 2 days \citep{she03}\footnote{It is worth noting here
that the combination of detections of correlated optical/X-ray variability
in different data sets for NGC~4051 makes the correlation extremely strong in this source, 
with the result that lags measured are quite robust, since only a single, well-sampled
event is needed in two light curves to measure an accurate lag, 
{\it assuming the correlation is real}.}.  
This result does not conflict with the 
{\it XMM-Newton} result however, which probes shorter time-scales.  Therefore, it is
possible that there are multiple time delays in the response of optical/UV to X-ray 
variability and vice versa, with a simple reprocessed component of optical/UV variability
on short time-scales, but some other component which leads X-ray variability on 
longer time-scales (e.g. due to propagation of accretion flow variations from the cooler, outer
disk to the inner X-ray emitting regions).

\subsection{NGC~3516}
\begin{figure}[!ht]
\plotfiddle{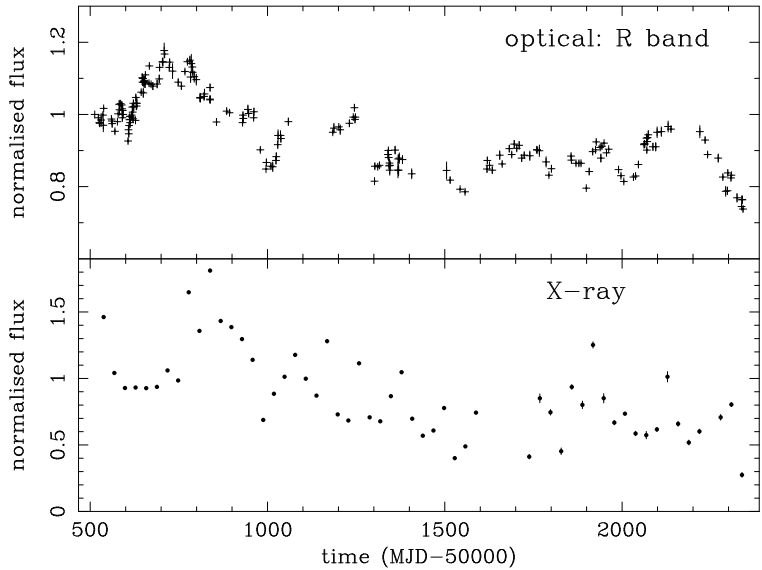}{8cm}{0}{60}{60}{-150}{-50}
\caption{NGC~3516 normalised R-band optical and 2-10~keV X-ray light curves (data taken
from \citealt{mao02}).} \label{ngc3516lcs}
\end{figure}

The Seyfert~1 NGC~3516 was monitored for 5 years by {\it RXTE} and in the optical
at the Wise Observatory.  Although an initial analysis of the first two years worth of data showed
tantalising evidence of a $\sim100$ day lag of X-rays to optical \citep{mao00}, 
the subsequent 3 years of observations showed no evidence for any correlation at any lag
(\citealt{mao02}, and see Fig.~\ref{ngc3516lcs}),
suggesting that the claimed correlation was due to low event statistics.
\citet{mao02} also showed that, unlike NGC~7469, NGC~3516 showed no evidence
for any correlation of X-ray spectral slope with optical flux, leaving the puzzling situation of
no apparent connection at all between optical and X-ray variability.  Furthermore, no
obvious relationship between the optical/UV and X-ray bands was seen
in variations on time-scales of a day or less (using simultaneous {\it HST}, {\it ASCA}
and {\it RXTE} data, \citealt{ede00}).  We note that there do appear to be monotonic
trends in the same direction in both optical and X-ray long term light curves, which could
be suggestive of a correlation on very long time-scales, 
but this could easily be due to chance.

\subsection{NGC~5548}
\begin{figure}[!ht]
\plotfiddle{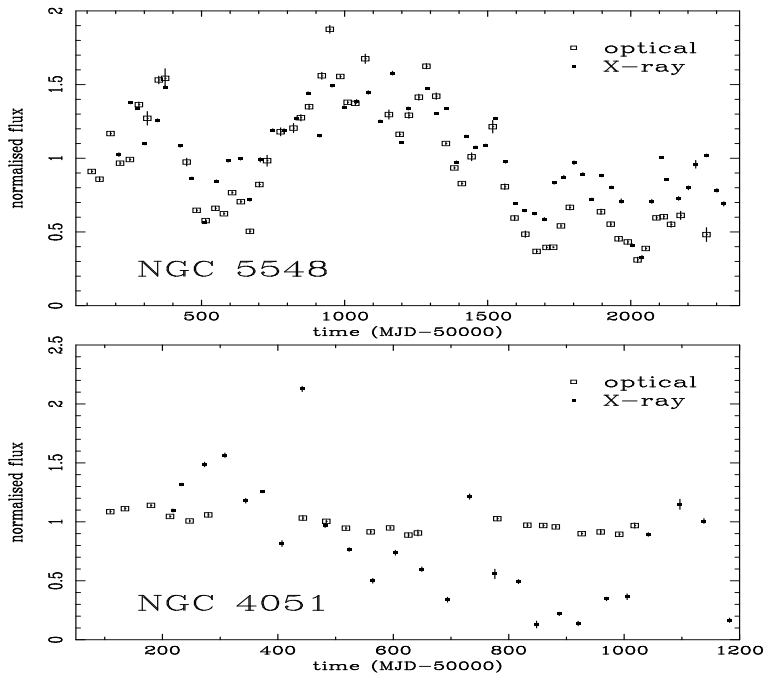}{8.5cm}{0}{90}{70}{-150}{-30}
\caption{Comparison of optical and X-ray light curves of NGC~5548 and NGC~4051.
The light curves have been binned up in 30~day bins, in order to smooth out the rapid X-ray
variability.  In NGC~5548 the amplitude of optical variability is quite large compared to X-rays, 
whereas in NGC~4051 the optical variability amplitude is much smaller than in X-rays,
even when accounting for the galaxy-bulge starlight contamination,
which contributes less than half of the
optical flux (e.g. \citealt{don90}).} \label{ngc5548_4051lcs}
\end{figure}

Arguably the best long-term simultaneous optical and X-ray light curves have been obtained
for the Seyfert~1 NGC~5548, with 6 years of overlapping data from the {\it RXTE} and 
AGN Watch monitoring campaigns.  These data showed the strongest correlation
yet observed between optical and X-ray variations (confirming the earlier suggestion
of a correlation by \citealt{cla92}), with no measurable lag down to
time-scales of a few weeks (see \citealt{utt03}, and Fig.~\ref{ngc5548_4051lcs}).  On short
time-scales (up to a few weeks), the X-rays are significantly more variable than
the optical band - this appears to be a general trend in all Seyfert galaxies monitored so far - 
although it isn't possible to say if the X-rays are still correlated with the optical on these
short time-scales in NGC~5548 (so we can't be sure we are dealing with the same component).

Interestingly however, the amplitude of optical variability in NGC~5548 is similar to the
amplitude of X-ray variability on longer time-scales of years, and when the 
contamination due to host galaxy starlight is subtracted, the amplitude of optical 
5100\AA\ variability
is even larger than in X-rays (43\% versus 31\% respectively).   This result seems to rule out 
the possibility that X-ray reprocessing drives the bulk of the optical variability, because
in that case the amplitude of optical variability should be at least equal to or less than
the amplitude of X-ray variability, as the optical variations due to reprocessing
might be diluted by emission due to internal heating.  Since the UV band varies even more
than the optical \citep{gil03}, the problem is even worse for reprocessing, 
although a possible (if ad hoc) explanation could be that we do not see the same X-ray
variations that the reprocessor sees, e.g. due to anisotropic X-ray emission.

\section{Putting the puzzle together}
\begin{figure}[!ht]
\plotone{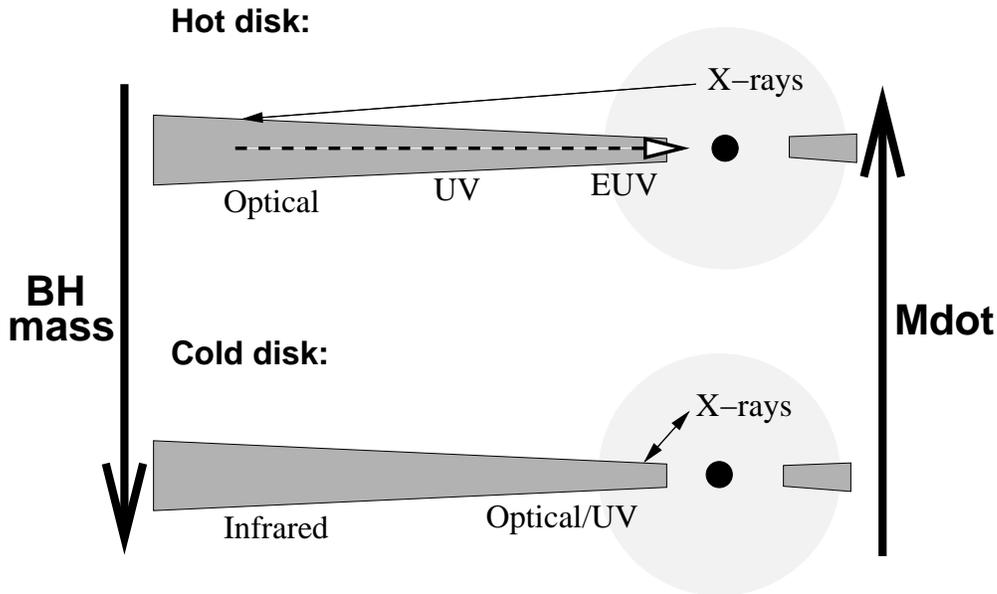}
\caption{Summary of some possible optical/X-ray interactions
in the `standard' AGN disk-corona picture, and how they depend on mass
and accretion rate.  Solid arrows represent interactions
due to photons, i.e. reprocessing and Comptonisation, while the dashed arrow 
denotes the indirect connection due to propagating accretion variations.} \label{toypic}
\end{figure}
The relations between optical and X-ray variability observed in Seyfert galaxies
present us with a puzzling picture.  Even though the sample observed so far is rather
limited, it is already clear that there are a variety of different behaviours.  The
four case studies presented here are also representative of the types of behaviour
seen in a few other Seyfert galaxies with simultaneous optical/X-ray observations, which we
have not detailed here.  Broadly speaking, we can discern three types of behaviour:
\begin{enumerate}
\item {\it Correlated optical and X-ray flux variability}: NGC~5548 and NGC~4051 fall into this 
category, as does the NLS1 Ark~564 \citep{she01}.  In all cases X-rays vary strongly, but
we find both weak optical variability (in NGC~4051 and Ark~564) and strong optical
variability (in NGC~5548).
\item {\it Correlated optical/UV flux and X-ray spectral variability}: NGC~7469 does not
show an obvious correlation between UV and 2-10~keV X-ray flux variations, but does show
a correlation between UV flux variations and changes in X-ray spectral slope.  This type of
correlation is probably not so common in Seyfert galaxies, as it implies no strong 
correlation between X-ray spectral slope and 2-10~keV flux (otherwise the X-ray flux would
be well-correlated with UV).  Most Seyferts seem to
show a strong correlation between spectral slope
and 2-10~keV flux however (e.g. \citealt{mar03b}).
\item {\it Uncorrelated optical and X-ray variability}: NGC~3516 shows no obvious correlation
between optical and X-ray flux variations, on either short or long time-scales, or
between optical flux variations and X-ray spectral variations.  This behaviour stands in 
sharp contrast to that of NGC~5548.
\end{enumerate}
The first class of behaviour, correlated flux variability, is what was originally anticipated,
and fits into a picture where the optical and X-ray variations are closely related, either through 
heating of the disk by X-rays, or Comptonisation of the optical seed photons so that X-rays
track the optical variations.  What is perhaps surprising is that the optical variability of NLS1,
such as NGC~4051 and Ark~564, is so weak compared to their X-ray variations, whereas
the amplitude of
optical variability of NGC~5548 is large, and is even greater than the X-ray variability amplitude,
at least on long time-scales.  A similar picture, of rather
weak optical variability in NLS1 is also presented
by \citet{kli04} (and see Gaskell, these proceedings).

However, this discrepancy might also be understood in terms of the `standard' picture of AGN. 
Fig.~\ref{toypic} demonstrates some of the possible ways that optical variations and X-ray 
variations might be interconnected.  The kinds of interaction we expect are likely to be
strongly governed by the disk temperature: in hotter disks, the optical emitting region is likely to
come from many gravitational radii
(e.g. $>1000$~R$_{\rm G}$ in NGC~4051)
from the central region where most of the accretion energy is liberated (and where the X-ray
emission 
likely originates).  Thus the main connection between X-ray and optical variations is likely to be
reprocessing, by X-ray heating of the optical-emitting disk at large radii (although
we also expect a correlation which has longer lags in the opposite direction, due to 
the propagation of variations in the accretion flow
as suggested by \citealt{she03}).  For the expected
geometry, the optical emitting region will subtend only a small fraction of the sky as seen
from the X-ray source, so it is likely that constant (at least
at large radii) internal heating will dominate the optical band, diluting the varaibility amplitude.
In cooler disks, the optical/UV emission will originate much closer to the X-ray emitting region
(and may be embedded within it), opening the way for other interactions, such as optical/UV 
variations (e.g. due to accretion instabilities in the inner disk) driving X-ray variations via 
Comptonisation.

The disk temperature is expected to be a function of black hole mass, $M_{\rm BH}$, and
accretion rate ($\dot{m}$ in Eddington units), scaling as $T\propto(\dot{m}/M_{\rm BH})
^{\frac{1}{4}}$, according to standard disk theory.  Since the emerging consensus is 
that NLS1 have relatively low black hole masses and are accreting at high rates (e.g.
see reviews by Boller, M$^{\rm c}$Hardy, these proceedings), it is perhaps not surprising
that they show rather weak optical variability.  On the other hand, the broad line Seyfert~1
NGC~5548 has a rather high mass ($\sim10^{8}$~M$_{\odot}$, \citealt{pet02}) and
low accretion rate, so its optical and X-ray emitting regions may be co-spatial, in which case
the larger amplitude of optical variability compared to X-rays may be because the 
optical is directly driving the X-ray variations through Comptonisation.

The other types of optical/X-ray relation, shown by NGC~7469 and NGC~3516, are more
difficult to explain in the standard picture.  \citet{pet04} have suggested geometrical changes
in the X-ray source as an explanation of the UV/X-ray-slope correlation (and lack of 
UV/X-ray-flux correlation) in NGC~7469.  In NGC~3516, the lack of any correlation is 
even more puzzling, and raises the possibility that the X-ray source and optical source are
somehow not `aware' of each other's variations, for example if the X-ray emission 
is anisotropic (e.g. see discussion by Gaskell, these proceedings).

Since other contributions at this meeting have demonstrated the importance of the analogy
with black hole X-ray binaries (BHXRBs) in understanding AGN behaviour (e.g. 
Uttley, M$^{\rm c}$Hardy, elsewhere in these proceedings), it might be fruitful to look
to BHXRBs for clues to the origin of different optical/X-ray relations.
In BHXRBs, the thermal disk emission is mainly seen in the X-rays, due to the much higher disk
temperatures ($kT\sim1$~keV) than in AGN.  In the disk-dominated `high/soft' state, the disk X-ray
emission is remarkably stable (e.g. \citealt{chu01}), and this result immediately suggests an
intriguing difference with AGN, perhaps because the inner disks of AGN are subject to
instabilities that are important at lower temperatures,
e.g. due to Hydrogen ionisation \citep{bur98}.  But on long time-scales, BHXRBs show evidence
for state transitions, where the strength of the corona and disk emission change on different
time-scales, suggesting two different accretion flows, a hot coronal flow, with short variability
time-scales, and a slower optically thick flow, i.e. the standard disk \citep{chu01}. 
It is interesting to speculate
whether the relative strengths of these components could play a role in the different
types of X-ray/optical relation.  For example, if optical and X-ray variations originate in 
two different accretion flows, with shorter variability time-scales (at a given radius) in the 
hot X-ray flow, then the optical and X-ray
variations we see on similar time-scales in NGC~3516 may originate
from two very different radii and hence be unconnected for that reason. 
Such a situation might occur in AGN where the energy release
in optical and X-ray emitting flows is evenly balanced, so neither dominates the variability in
the other band (e.g. through reprocessing or Comptonisation).

Much progress has already been made, simply by observing a few objects.
However, in order to make further progress in disentangling the complex connections
between optical and X-ray emission components in AGN, simultaneous optical/X-ray monitoring of
a larger sample (including quasars) is required, which covers a wide range of
black hole mass and accretion rate.  The new generation of robotic or queue-scheduled optical
telescopes will help to facilitate such campaigns, but a continued, more sensitive
X-ray monitoring capability is urgently needed to pick up the reins from {\it RXTE}
when it reaches the end of its remarkably productive life.

\end{document}